\documentclass[journal]{IEEEtran}
\usepackage[english]{babel}
\usepackage{amsmath}
\usepackage{graphicx}
\usepackage[colorlinks=true, allcolors=blue]{hyperref}
\usepackage{xcolor}
\usepackage{comment}
\usepackage{lipsum}
\usepackage{lettrine}
\usepackage{cite}  
\usepackage{multicol}
\usepackage{makecell}
\usepackage{colortbl}
\usepackage{xcolor}
\usepackage{booktabs}
\usepackage{balance}
\usepackage{csquotes}
\usepackage{adjustbox}
\usepackage[caption=false,font=normalsize,labelfont=sf,textfont=sf]{subfig}
\usepackage{placeins}
\usepackage{float}
\usepackage{amssymb}

\newcommand{\orcidiconBeo}{\href{https://orcid.org/0000-0001-5929-1672}{\includegraphics[scale=0.1]{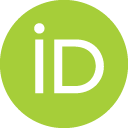}}}

\newcommand{\orcidiconOba}{\href{https://orcid.org/0000-0003-2523-3858}{\includegraphics[scale=0.1]{figures_eps/orcidID128.png}}}

\begin{document}
\bstctlcite{IEEEexample:BSTcontrol}
\setlength{\parskip}{0pt}

\title{Gut-Brain Axis as a Closed-Loop Molecular Communication Network}

\author{Beyza E. Ortlek\orcidiconBeo,~\IEEEmembership{Student Member,~IEEE}, and~Ozgur~B.~Akan\orcidiconOba,~\IEEEmembership{Fellow,~IEEE}
\thanks{Beyza E. Ortlek is with the Center for neXt-generation Communications (CXC), Department of Electrical and Electronics Engineering, Ko\c{c} University, Istanbul 34450, Turkey (e-mail: bortlek14@ku.edu.tr).}
\thanks{Ozgur B. Akan is with the Center for neXt-generation Communications (CXC), Department of Electrical and Electronics Engineering, Ko\c{c} University, Istanbul 34450, Turkey and also with the Internet of Everything (IoE) Group, Electrical Engineering Division, Department of Engineering, University of Cambridge, Cambridge CB3 0FA, UK (email: oba21@cam.ac.uk).}
\thanks{This work was supported in part by the AXA Research Fund (AXA Chair for Internet of Everything at Ko\c{c} University).}
}

\maketitle

\begin{abstract}
\label{sec:abstract}
Molecular communication (MC) provides a quantitative framework for analyzing information transfer within biological systems. This paper introduces a novel and comprehensive MC framework for the gut-brain axis (GBA) as a system of six coupled, nonlinear delay differential equations (DDEs). The proposed model defines a bidirectional feedback loop with a gut-to-brain inflammatory channel and a brain-to-gut neuroendocrine channel. Under prolonged stress, this feedback loop becomes self-perpetuating and drives the system into a pathological state. We evaluate the end-to-end channel across varying conditions using time-domain simulations, small-signal frequency-domain characterization, and an information-theoretic capacity analysis. At homeostasis, the system maintains stable circadian dynamics with higher information throughput, whereas sustained stress drives a shift to dysregulated hypercortisolism. In this pathological state, spectral efficiency decreases due to a narrowed effective bandwidth and a lower passband gain driven by neuroendocrine delays and saturating cytokine–hormone kinetics. These results quantify the impact of these signaling mechanisms on stability and information processing, elucidating the transition from healthy circadian rhythms to a persistent pathological state of hypercortisolism.
\end{abstract}


\section{Introduction}
\label{sec:intro}

\IEEEPARstart{M}{olecular} communication (MC) is a bio-inspired paradigm that utilizes molecules as information carriers, establishing a quantitative framework to analyze biological signaling networks from an information and communication theoretical perspective \cite{akan2016fundamentals,farsad2016comprehensive}. By modeling the transmission, reception, and propagation of signaling molecules, MC provides the theoretical tools to decipher complex biological processes, contrasting with conventional electromagnetic models that cannot capture the stochastic, diffusion-based nature of these systems. The MC paradigm is foundational to the emerging concept of the Internet of Bio Nano Things (IoBNT), which envisions the integration of biological and artificial nanoscale entities into a communication network for novel applications in healthcare, diagnostics, and personalized medicine \cite{kuscu2021internet, akyildiz2015internet}.

The human body operates as a highly complex biological system, relying on intricate signaling networks to maintain physiological homeostasis. Understanding the communication principles that govern these networks is critical for deciphering the mechanisms underlying health and disease. MC provides the necessary analytical framework to investigate these signaling networks. A quintessential example of such a network is the gut-brain axis (GBA), which is a bidirectional communication system linking the gastrointestinal tract and the central nervous system (CNS) \cite{cryan2019microbiota, mayer2011gut}. This axis integrates the gut and its vast microbial community with the brain, forming a complex information network that influences a wide range of physiological and psychological processes \cite{cryan2012mind, clarke2014minireview}.

\begin{figure*}[!t]
	\centering
	\includegraphics[width=1\textwidth]{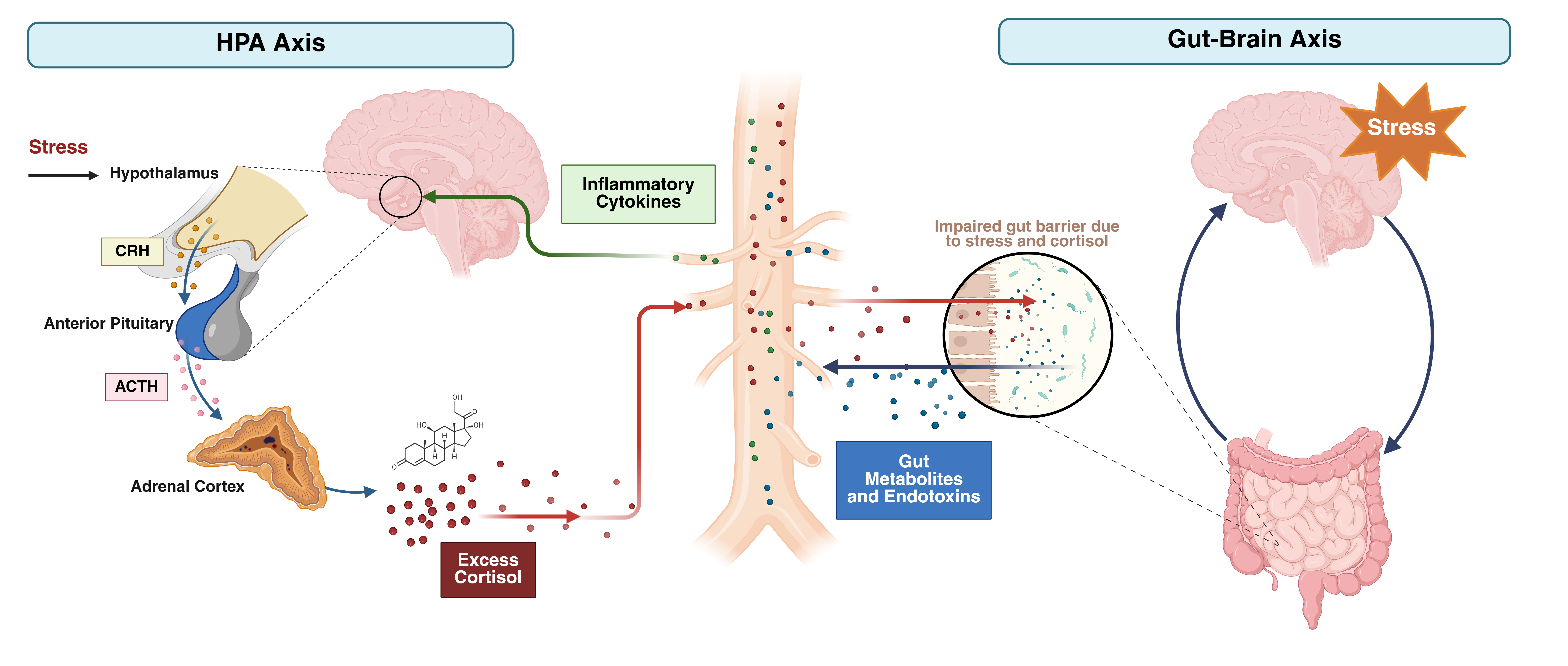}%
	\caption{Illustration of bidirectional communication along the gut–brain axis via the HPA axis. (Created in  https://BioRender.com).}
    \label{fig:bidirectional}%
\end{figure*}

The intricate communication within the GBA is facilitated by intertwined neural, immune, and endocrine channels that create a robust and dynamic network that is essential for homeostasis \cite{grenham2011brain, ding2017gut}. The neural pathway provides rapid signaling through the vagus nerve and the enteric nervous system (ENS) \cite{bonaz2018vagus, cryan2019microbiota, ortlek2025modeling}. The endocrine channel uses circulating neuroactive compounds often produced by the gut microbiota, while the immune channel relies on signaling molecules called cytokines to communicate with the brain \cite{clarke2014minireview, el2014immune}. A principal neuroendocrine component of the GBA is the HPA axis, which is responsible for the core stress response system of the human body. The HPA axis operates as a closed-loop system where bottom-up signals from the gut, such as pro-inflammatory cytokines, trigger top-down neuroendocrine responses from the brain. These responses, in turn, modulate the gut environment by altering motility, secretion, and intestinal permeability, establishing a feedback cycle critical for homeostasis \cite{bertollo2025hypothalamus, mayer2011gut}. As illustrated in Fig. \ref{fig:bidirectional}, the integrity of this cycle is fundamentally dependent on the intestinal barrier, where increased permeability allows microbial-associated molecular patterns, like lipopolysaccharide (LPS), to translocate from the gut into the circulation. This translocation initiates a systemic inflammatory response that drives further bottom-up signaling through both humoral and neural routes, perpetuating the cycle and directly impacting HPA axis regulation \cite{cryan2019microbiota}.

Dysregulation of signaling within the GBA is implicated in a broad spectrum of disorders, with strong links to stress-related physiological and psychological conditions \cite{cryan2019microbiota, grenham2011brain}. Both acute and chronic stress disrupt homeostasis by altering the composition of the gut microbiota, a state known as dysbiosis. This increases intestinal permeability, allowing pro-inflammatory molecules like LPS to translocate from the gut into the circulation, which triggers a systemic immune response \cite{kasarello2023communication}. This inflammatory signal activates the HPA axis, the body's primary stress response system. In a healthy individual, the HPA axis follows a cascade where corticotropin-releasing hormone (CRH) from the hypothalamus triggers the pituitary to release adrenocorticotropic hormone (ACTH), which causes the adrenal cortex to release cortisol. This cortisol then executes a negative feedback loop to restore homeostasis. However, under the persistent stimulation of chronic stress, this crucial regulatory feedback becomes impaired \cite{bertollo2025hypothalamus, dinan2006hypothalamic,malek2015dynamics}.

The persistent activation of the HPA axis leads to elevated levels of stress hormones like cortisol, which further weaken the gut barrier. This compromised barrier permits increased translocation of microbial components such as LPS into the bloodstream. The subsequent systemic immune response triggers the release of pro-inflammatory cytokines like interleukin 6 (IL-6) and tumor necrosis factor alpha (TNF-$\alpha$), which further stimulates the HPA axis. This feedback loop sustains a state of low-grade systemic inflammation coupled with neuroendocrine dysregulation, a condition strongly implicated in the pathophysiology of mood disorders like depression and anxiety \cite{cryan2012mind,tofani2025gut}. The resulting state of chronic hypercortisolism impairs the HPA axis's negative feedback, leading to sustained hormone production and neurotoxic effects in brain regions critical for mood regulation, such as the hippocampus and prefrontal cortex. Evidence from both germ-free animal and clinical studies supports a causal role for the microbiota in this process, strengthening the motivation for a unified communication framework that models these microbial and host interactions \cite{cryan2019microbiota, mayer2011gut}.

Given the complexity of the GBA, MC modeling is crucial for quantitatively analyzing its signaling dynamics and uncovering the mechanisms that contribute to disease. Traditional systems-level frameworks often neglect critical features of biological signaling, such as molecular noise, spatial diffusion, and temporal delays. In contrast, MC provides a comprehensive and detailed modeling approach that intrinsically incorporates these dynamics through advection-diffusion and ligand-receptor interactions. This makes it a uniquely suitable methodology for quantitatively analyzing the intricate pathways of the GBA and deciphering how disruptions in these signaling pathways can lead to pathological states.

Existing research has established the foundation for analyzing the GBA using the MC approaches. Foundational work introduced the Microbiome-Gut-Brain Axis as a comprehensive biomolecular communication network, outlining the potential for IoBNT integration \cite{akyildiz2019microbiome}. Subsequent studies have developed end-to-end MC models for specific gut-to-brain routes, including the signaling of the microbial metabolite p-cresol in the context of Autism Spectrum Disorder \cite{ortlek2023communication} and SCFA-driven vagus nerve signaling \cite{ortlek2025modeling}. Concurrently, other research has explored the GBA at the synaptic scale, for instance by modeling neurotransmitter diffusion as a Molecular Quantum (MolQ) communication channel \cite{maitra2024molecular}. Collectively, these studies have clarified how molecular messages are produced, transported, and decoded along individual pathways and have demonstrated the value of communication-theoretic tools in this domain. However, the focus has remained mainly on unidirectional signaling or on isolated components of the axis, which limits insight into the system-level behavior that emerges when multiple channels interact within a feedback structure.

In this work, we address this gap by proposing a novel MC framework that models the GBA as a bidirectional, closed-loop network. While prior mathematical studies have coupled inflammatory cytokines to the HPA axis \cite{malek2015dynamics}, a complete representation that closes the loop via the brain-to-gut pathway with its inherent time delays is still lacking. Therefore, we formulate a system of delay differential equations (DDEs) that explicitly links HPA axis activity to gut barrier integrity, thereby closing the communication loop. Utilizing the proposed model, we simulate the system's response to healthy, acute, and chronic stress profiles to characterize the transition from homeostasis to a pathological state of hypercortisolism. Furthermore, we quantify the performance of communication channel performance by analyzing its stability, small-signal gain, and information-theoretic capacity, identifying key parameters that govern the system's dynamics. This approach moves beyond the analysis of individual pathways to provide a holistic, systems-level understanding of GBA dynamics in both health and disease. By conceptualizing this network as a communication channel, the study aims to:

\begin{itemize}
    \item Formulate a system of delay differential equations (DDEs) to model the closed-loop dynamics between gut permeability, circulating endotoxin, pro-inflammatory cytokines (TNF-$\alpha$, IL-6), and HPA axis hormones (ACTH, Cortisol).
    \item Simulate the response of the system to different stress profiles (healthy, acute, chronic) to characterize the transition from healthy circadian oscillations to a pathological state of hypercortisolism.
    \item Analyze the GBA as a communication channel by linearizing the model to derive its end-to-end transfer function.
    \item Quantify the channel's performance by calculating key communication metrics, including small-signal gain, bandwidth, and information-theoretic capacity, to understand how chronic stress impairs the GBA's ability to process information.
\end{itemize}

The rest of the paper is organized as follows. Section~\ref{sec:system_model} presents the bidirectional closed-loop molecular communication model and defines the six state variables, delay terms, and channel couplings. Section~\ref{sec:theoretical_analysis} reports the theoretical analysis and simulations, including healthy, acute, and chronic stress scenarios. It presents an analysis to identify stability thresholds and a frequency-domain analysis to quantify the channel's small-signal gain, bandwidth, and information-theoretic capacity. Section~\ref{sec:discussion} interprets the results in the context of gut barrier control and hypothalamic-pituitary-adrenal regulation and identifies potential therapeutic targets. Section~\ref{sec:conclusion} summarizes the contributions, outlines limitations, and suggests directions for future work.

\section{System Model} 
\label{sec:system_model}

The GBA is a sophisticated, bidirectional communication network that is crucial for maintaining physiological and psychological homeostasis. The proposed framework models a vital feedback loop within this communication network, detailing the signaling mechanism between the HPA axis, immune activation, and gut barrier integrity. Under chronic stress conditions, this regulatory system transforms into a self-perpetuating process of inflammation and neuroendocrine dysregulation. The proposed model is developed to quantitatively examine these dynamics as a system of six coupled, nonlinear DDEs. The communication system is characterized by six primary signals, representing the concentrations or levels of endotoxin ($P(t)$), TNF-$\alpha$ ($T(t)$), IL-6 ($S(t)$), ACTH ($A(t)$), cortisol ($C(t)$), and gut permeability ($L(t)$).

\subsection{Model Architecture and Signal Propagation}

The model architecture is built upon three fundamental feedback loops that describe the communication network:

\begin{enumerate}
    \item \textbf{The HPA Negative Feedback Loop:} Cortisol provides delayed ($\tau_{\text{hpa}}$) negative feedback to inhibit ACTH and its own production, depicting the canonical endocrine regulatory mechanism.
    \item \textbf{The Gut-to-Brain Inflammatory Pathway:} Systemic inflammation, driven by endotoxin ($P$) leaking from a compromised gut barrier, stimulates the HPA axis. Cytokines ($T$ and $S$) promote the production of HPA hormones, i.e., ACTH and cortisol.
    \item \textbf{The Brain-to-Gut Damage Pathway:} Elevated levels of cortisol increase gut permeability ($L$), leading to further endotoxin leakage via a delayed feedback mechanism ($\tau_{\text{gut}}$).
\end{enumerate}

\begin{figure*}[!t]
	\centering
	\includegraphics[width=1\textwidth]{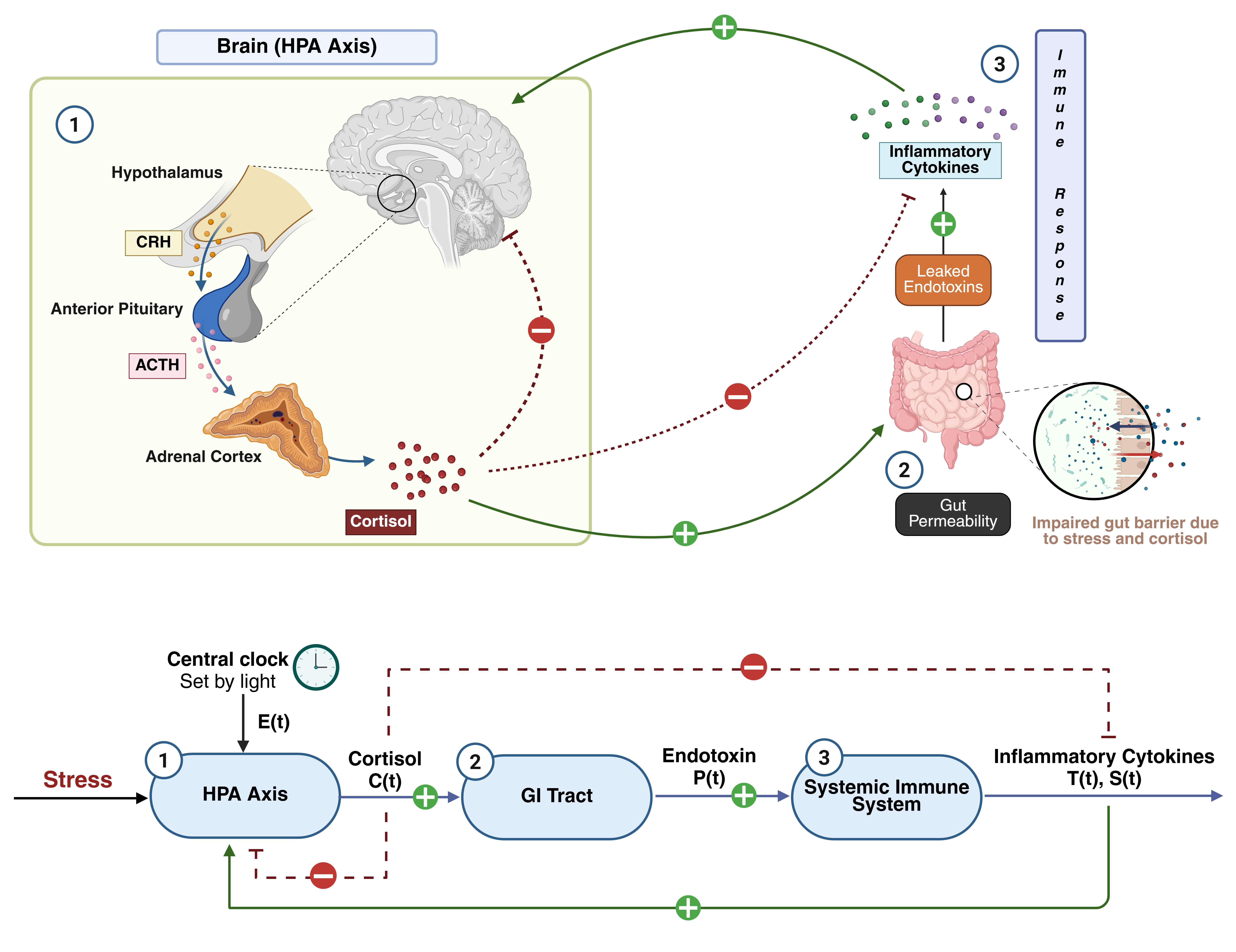}%
	\caption{Closed-loop communication within the gut–brain axis via the HPA axis. (Created in  https://BioRender.com).}
	\label{fig:feedback_mc}
\end{figure*}

These feedback loops initiate a self-perpetuating cycle under chronic stress conditions. The process begins when elevated cortisol ($C$) increases gut permeability ($L$) after a delay ($\tau_{\text{gut}}$), allowing endotoxins ($P$) to enter the bloodstream. This triggers an immune response and the release of pro-inflammatory cytokines ($T$ and $S$), which in turn transmit a powerful inflammatory signal back to the brain, stimulating the HPA axis to produce even more cortisol ($C$). This resulting surge of cortisol closes the loop by further damaging the gut barrier, thus reinforcing and sustaining the entire pathological cycle.

\subsection{Molecular Communication Framework}

Fig.~\ref{fig:feedback_mc} provides a schematic overview of the proposed closed-loop MC model. Within this MC framework, the system can be conceptualized as follows:

\begin{itemize}
	\item \textbf{Input Signal}: An external stressor, which is modeled as an increase in the leakage rate of endotoxin molecules from the gut. This represents the initial trigger that perturbs the system from its healthy state.
	\item \textbf{Channel and Transmission Mechanism}:  The entire bidirectional physiological cascade, encompassing the hormonal brain-to-gut pathway, the endotoxin-driven gut-to-immune pathway, and the cytokine-mediated immune-to-brain feedback loop.
	\item \textbf{Output Signal}: The plasma concentration of cortisol, $C(t)$, which reflects the system's overall response to the input stressor.
\end{itemize}

The dynamics of the proposed model are described by the rate of change for six key state variables ($L(t)$, $P(t)$, $T(t)$, $S(t)$, $A(t)$, $C(t)$). The following equations form a coupled system in which the output of one physiological process serves as the input to another. This interconnected structure mathematically defines the vital feedback loops that govern GBA regulation. The communication process is initiated by the \textbf{Brain $\rightarrow$ Gut} pathway, where the final output of the HPA axis, cortisol, modulates the integrity of the intestinal barrier. The state of gut permeability, $L(t)$, is modeled as a dynamic balance between cortisol-induced damage and the gut's natural repair mechanisms:

\begin{equation}
\begin{split}
    \frac{dL(t)}{dt} &= \underbrace{k_{\text{damage}} \left( \frac{C(t-\tau_{\text{gut}})^{n_{\text{gut}}}}{C_{\text{half}}^{n_{\text{gut}}} + C(t-\tau_{\text{gut}})^{n_{\text{gut}}}} \right)}_{\text{Damage from Cortisol}} \\ \\
    &- \underbrace{k_{\text{repair}} (L(t) - L_{\text{base}})}_{\text{Natural Repair}},
\end{split}
\end{equation}

\noindent where the first term models the damage inflicted by past cortisol levels, $C(t-\tau_{\text{gut}})$, using a Hill function. This captures a nonlinear saturating effect where the damage rate increases with cortisol concentration but eventually plateaus. The time delay $\tau_{\text{gut}}$ accounts for hormonal circulation and cellular response time. The second term represents the gut's intrinsic capacity for self-repair, modeled as a linear process that drives the permeability level $L(t)$ back toward its healthy baseline $L_{\text{base}}$.

A compromised gut barrier allows endotoxin molecules to pass from the intestinal lumen into the bloodstream, initiating the \textbf{Gut $\rightarrow$ Immune} signaling pathway. The dynamics of the resulting plasma endotoxin concentration, $P(t)$, which serves as the input signal for the inflammation process, are described by:

\begin{equation}
\begin{split}
    \frac{dP(t)}{dt} &= \underbrace{k_{\text{leak}} L(t)}_{\text{Leakage from Gut}} - \underbrace{e_P P(t)}_{\text{Natural Clearance}} \\ \\
    &- \underbrace{d_1 \left( \frac{T(t)}{x_1 + T(t)} \right) \left( \frac{S(t)}{x_2 + S(t)} \right) P(t)}_{\text{Immune-Mediated Clearance}} .
\end{split}
\end{equation}

The production of endotoxin is proportional to the gut permeability ($L$) and modulated by the leakage rate $k_{\text{leak}}$, which is a result of the stress input. Endotoxin is subsequently cleared from circulation via a natural, first-order elimination process at a rate $e_P$, and also through a more potent clearance by the immune system, which is dependent on the pro-inflammatory cytokines TNF-$\alpha$ ($T$) and IL-6 ($S$).

The presence of endotoxin in the bloodstream triggers a pro-inflammatory cytokine cascade, starting with the production of TNF-$\alpha$, whose dynamics are modeled as:

\begin{equation}
\begin{split}
    \frac{dT(t)}{dt} &= -e_T T(t) + \underbrace{k \frac{P(t)}{x_3 + P(t)}}_{\text{Stimulation by LPS}} \\ \\ 
    &+ \underbrace{\left( \frac{d_2}{x_4 + C(t)} + \frac{d_3}{x_5 + S(t)} \right) \left( \frac{T(t)}{x_6 + T(t)} \right)}_{\text{Self-Stimulation (Inhibited by Cortisol \& IL-6)}} .
\end{split}
\end{equation}

The production of TNF-$\alpha$ is driven by endotoxin ($P$), a process modeled with Michaelis-Menten kinetics to capture saturation effects. The equation also incorporates a positive feedback term where TNF-$\alpha$ stimulates its production to amplify the immune signal rapidly. This self-amplification is tightly regulated, suppressed by the anti-inflammatory effects of both cortisol ($C$) and the secondary cytokine IL-6 ($S$), providing a crucial control mechanism.

TNF-$\alpha$ then stimulates the production of IL-6, a secondary cytokine with both pro- and anti-inflammatory roles. The dynamics of $S(t)$ are given by:

\begin{equation}
\frac{dS(t)}{dt} = -e_S S(t) + \underbrace{d_4 \left( \frac{1}{x_7 + C(t)} \right) \left( \frac{T(t)}{x_8 + T(t)} \right)}_{\text{Stimulation by TNF-}\alpha\text{ (inhibited by cortisol)}} .
\end{equation}

This equation shows that IL-6 is produced in response to TNF-$\alpha$ levels. The synthesis of IL-6 is strongly inhibited by cortisol, highlighting cortisol's systemic role in modulating and ultimately suppressing the inflammatory cascade.

The circulating cytokines transmit an inflammatory signal to the brain, representing the \textbf{Gut $\rightarrow$ Brain} pathway, which directly influences the HPA axis. The dynamics of the ACTH ($A$) model the pituitary's response, which is driven by the body's internal clock and feedback from both cortisol and the immune system:

\begin{equation}
\begin{split}
    \frac{dA(t)}{dt} &= -e_A A(t) + \underbrace{(h E(t)) \frac{c^{m_1}}{c^{m_1} + C(t-\tau_{\text{hpa}})^{m_1}}}_{\text{Circadian Drive and Cortisol Inhibition}} \\ \\ 
    &+ \underbrace{d_5 \left( \frac{S(t)}{x_9 + S(t)} \right) \left( \frac{T(t)}{x_{10} + T(t)} \right)}_{\text{Stimulation by Cytokines}}.
\end{split}
\end{equation}

ACTH ($A$) production is driven by a baseline stimulus $h$ and modulated by a 24-hour circadian rhythm $E(t)$. This production is inhibited by delayed negative feedback from cortisol, $C(t-\tau_{\text{hpa}})$, and is stimulated by the inflammatory cytokines $S$ and $T$, which convey the molecular alarm signals from the periphery to CNS.

Finally, cortisol ($C$) is the terminal output of the HPA axis, and its production completes the multiple feedback loops within the communication system. The cortisol dynamics are described by:

\begin{equation}
\begin{split}
    \frac{dC(t)}{dt} &= -e_C C(t) + \underbrace{\alpha \frac{A(t-\tau_{\text{hpa}})^{m_2}}{a^{m_2} + A(t-\tau_{\text{hpa}})^{m_2}}}_{\text{Stimulation by ACTH}} \\ \\ 
    &+ \underbrace{d_6 \left( \frac{S(t)}{x_{11} + S(t)} \right) \left( \frac{T(t)}{x_{12} + T(t)} \right)}_{\text{Stimulation by Cytokines}}.
\end{split}
\end{equation}

Cortisol ($C$) is produced in response to past levels of ACTH ($A(t-\tau_{\text{hpa}})$). The inflammatory cytokines ($S$ and $T$) also directly stimulate the cortisol production, providing a parallel pathway for the immune system to amplify the stress response. The produced cortisol then acts back on the gut and the immune cells, closing the overall gut-brain communication loop.

\begin{table}[h!]
\centering
\caption{Model Parameters}
\label{tab:parameters_condensed}
\begin{tabular}{@{}p{1.1cm} p{4.3cm} p{2.8cm}@{}} 
\toprule
\textbf{Parameter}      & \textbf{Description}       & \textbf{Value} \\ \midrule
\multicolumn{3}{l}{\textbf{HPA Axis}} \\ \midrule
$h$       & Baseline for ACTH production               & 7.66 (pg/mL)/min \cite{malek2015dynamics} \\
$c$       & Half-saturation for cortisol inhibition    & 6.11 $\mu$g/dL \cite{malek2015dynamics} \\
$m_1$     & Hill coefficient for cortisol inhibition   & 4 (dimensionless) \cite{malek2015dynamics} \\
$\alpha$  & Max production rate of cortisol            & 0.28 ($\mu$g/dL)/min \cite{malek2015dynamics} \\
$a$       & Half-saturation for ACTH stimulation       & 21 pg/mL \cite{malek2015dynamics} \\
$m_2$     & Hill coefficient for ACTH stimulation      & 4 (dimensionless) \cite{malek2015dynamics} \\
$e_A$     & Elimination rate of ACTH                   & 0.04 min$^{-1}$ \cite{malek2015dynamics} \\
$e_C$     & Elimination rate of Cortisol               & 0.01 min$^{-1}$ \cite{malek2015dynamics} \\
$\tau_{\text{hpa}}$   & Time delay for HPA axis        & 10 min \cite{malek2015dynamics} \\
\midrule
\multicolumn{3}{l}{\textbf{Immune System}} \\ \midrule
$k$    & Production rate of TNF-$\alpha$ from LPS      & 0.0504 (ng/cL)/min \cite{malek2015dynamics} \\
$e_P$  & Elimination rate of LPS                       & 0.05 min$^{-1}$ \cite{malek2015dynamics} \\
$e_T$  & Elimination rate of TNF-$\alpha$              & 0.038 min$^{-1}$ \cite{malek2015dynamics} \\
$e_S$  & Elimination rate of IL-6                      & 0.02 min$^{-1}$ \cite{malek2015dynamics} \\ 
\midrule
\multicolumn{3}{l}{\textbf{Gut Permeability}} \\ \midrule
$k_{\text{damage}}$   & Rate of cortisol-induced gut damage      & 0.002 min$^{-1}$  \\
$k_{\text{repair}}$   & Natural gut repair rate                  & 0.05 min$^{-1}$  \\
$k_{\text{leak}}$     & Time-varying LPS leakage rate            & scenario-specific (min$^{-1}$) \\
$L_{\text{base}}$     & Baseline healthy gut permeability        & 0.1 (dimensionless)  \\
$C_{\text{half}}$     & Cortisol half-saturation                 & 15 $\mu$g/dL  \\
$n_{\text{gut}}$      & Hill coefficient                         & 2 (dimensionless)  \\
$\tau_{\text{gut}}$   & Time delay for cortisol                  & 120 min  \\
\bottomrule
\end{tabular}
\end{table}

\begin{figure*}[!t]
	\centering
	\includegraphics[width=\textwidth]{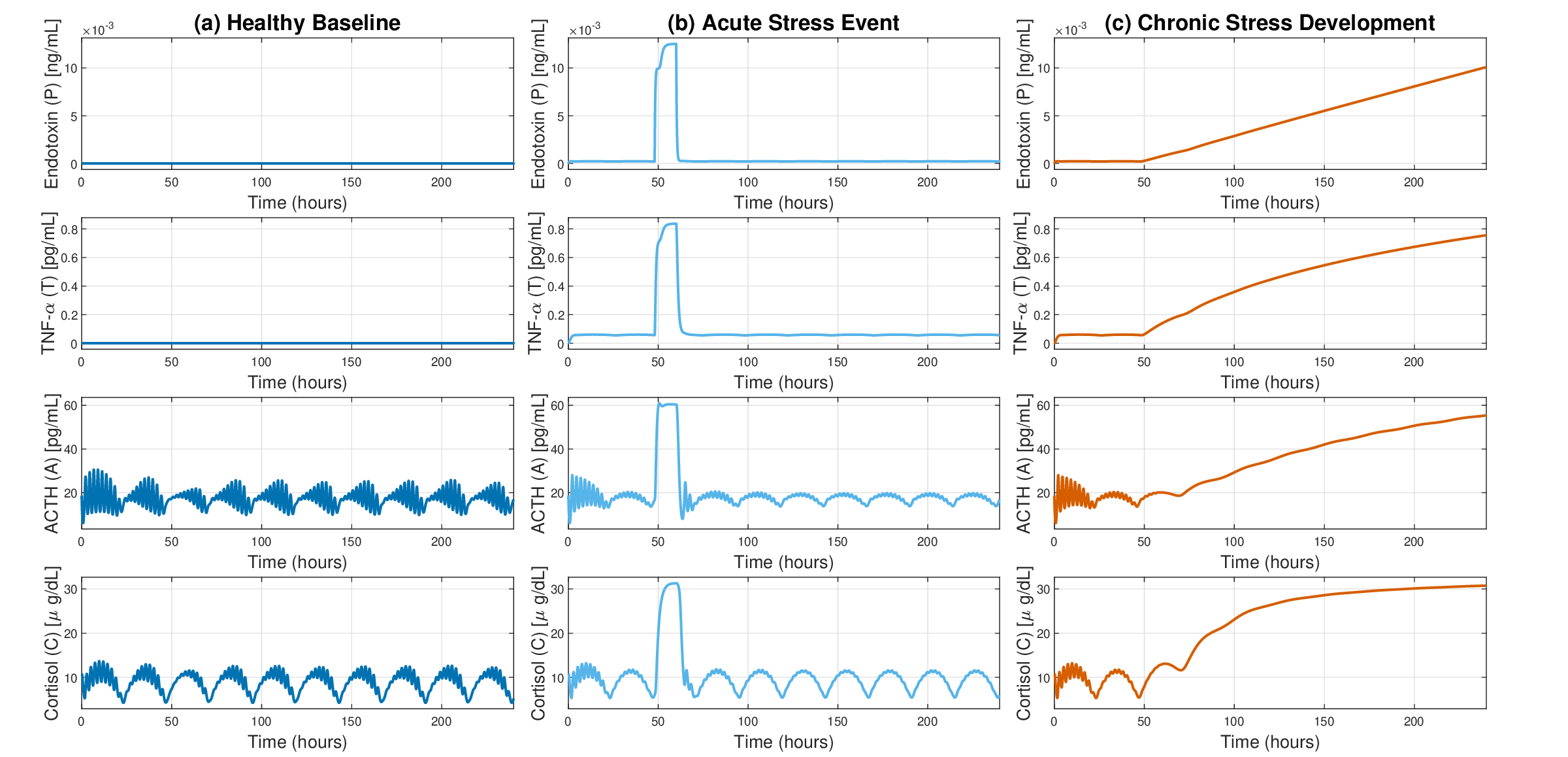}
    \caption{Comparative dynamics of the GBA network under three simulated stress scenarios. The columns depict the system's response to (a) a Healthy Baseline, (b) an Acute Stress event (a 12-hour pulse starting at hour 48), and (c) a Chronic Stress profile (initiated at hour 48). The rows display the time-course for the key signaling molecules: Endotoxin, TNF-$\alpha$, ACTH, and Cortisol.}
	\label{fig:stress_scenarios}
\end{figure*}

The system of coupled DDEs, with parameters specified in Table~\ref{tab:parameters_condensed}, forms the mathematical basis of our molecular communication framework for the closed-loop GBA. The parameter values for the HPA axis and immune components are adopted from the validated model by Malek et al. \cite{malek2015dynamics}, ensuring a physiologically grounded core. The novel gut permeability parameters were estimated to align with established physiological behavior. This MC-based model systematically represents the key processes underlying the stress-induced pathological cycle by integrating the HPA axis, peripheral immune signaling, and gut barrier dynamics. Although this framework does not aim to capture all the biological complexities of the GBA, it highlights the critical signaling mechanisms and feedback loops that govern system stability. Therefore, the proposed end-to-end model offers a structured approach to quantitatively analyze how the GBA responds to stress, providing a foundational basis for the theoretical analysis and the following numerical simulations.

\section{Theoretical Analysis and Simulations} 
\label{sec:theoretical_analysis}

To investigate the dynamics of the proposed closed-loop GBA network, numerical simulations of the system of DDEs defined in Section~\ref{sec:system_model} are conducted. The primary objective of this analysis is to characterize the system's transition from a healthy, homeostatic state to a pathological, dysregulated state under different stress conditions. The DDE system was solved numerically in MATLAB using the \texttt{dde23} solver. The end-to-end system model is simulated under three distinct scenarios, each initiated by a different external stress input profile. The results, shown in Fig.~\ref{fig:stress_scenarios}, display the system's temporal response over a 10-day period.

\subsection{Time-Domain Response to Stress Scenarios}

\paragraph{Healthy Baseline}
In the healthy baseline scenario, the system represents a non-stressed state governed by its internal regulatory signaling mechanisms.  As depicted in Fig.~\ref{fig:stress_scenarios}(a), the HPA axis operates under its canonical negative feedback, where cortisol production maintains stable circadian oscillations. Consequently, the effect of cortisol on the intestinal barrier is negligible, and gut permeability ($L$) is kept at a low baseline by natural repair processes. This ensures that the gut-to-brain inflammatory pathway, depicted in Fig.~\ref{fig:feedback_mc}, remains quiescent. Circulating levels of endotoxin molecules and the pro-inflammatory cytokine TNF-$\alpha$ are therefore kept at basal concentrations, reflecting an intact gut barrier and an inactive immune state. This scenario demonstrates that in the absence of significant stressors, the GBA network successfully maintains homeostasis.

\paragraph{Acute Stress} 
An acute stress event is modeled as a transient, 12-hour increase in the gut leakage rate $k_{\text{leak}}$ starting at hour 48. This perturbation raises circulating endotoxin, which triggers a rapid cytokine response and, in turn, stimulates the HPA axis. The HPA axis activation causes a surge in ACTH and cortisol that overrides the normal circadian rhythm, as seen in Fig.~\ref{fig:stress_scenarios}(b). After the $\tau_{\text{gut}}$ delay, this elevated cortisol increases gut permeability, leading to a temporary dysfunction of the gut barrier.

\begin{figure}[!h]
	\centering
	\includegraphics[width=\columnwidth]{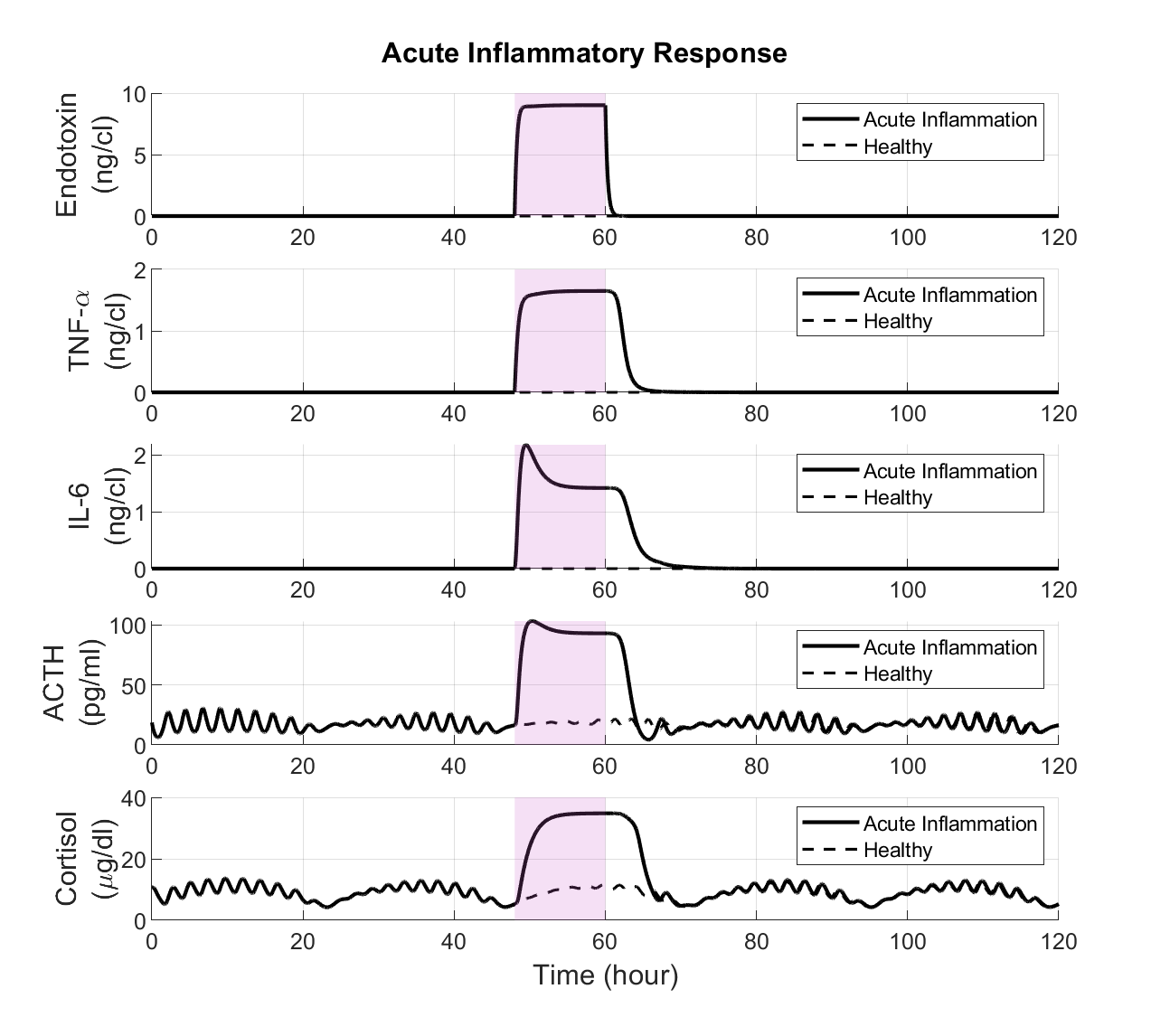}
    \caption{Dynamics of the acute inflammatory cascade following a transient endotoxin leakage.}
	\label{fig:acute_inf}
\end{figure}

The resulting temporary gut barrier dysfunction allows a further influx of endotoxin into systemic circulation, which activates the gut-to-brain inflammatory communication channel. This acts as a potent molecular signal, initiating a transient but robust inflammatory cascade characterized by a sharp rise in TNF-$\alpha$ and subsequently IL-6 as illustrated in Fig.~\ref{fig:acute_inf}. These circulating cytokine molecules constitute a secondary, delayed stimulatory signal that propagates back to the HPA axis, amplifying the initial cortisol peak and closing the bidirectional communication loop depicted in Fig.~\ref{fig:feedback_mc}. Once the external stressor is removed, the hyperactivation of the HPA axis ceases, resulting in a fall in cortisol levels. This facilitates the repair of the gut barrier, which terminates the endotoxin leakage and the subsequent inflammatory feedback signal. Within approximately 48 hours (Fig.~\ref{fig:acute_inf}), the system returns to its homeostatic circadian rhythm, underlining its capacity to manage and recover from acute physiological challenges effectively.

\paragraph{Chronic Stress Scenario}
Chronic stress is modeled as a sustained elevation of $k_{\text{leak}}$, which causes persistent activation of the HPA axis. This continuous primary input leads to chronically elevated cortisol levels, representing a dysregulated neuroendocrine output. This signal propagates down the brain-to-gut pathway, where the sustained high cortisol inflicts continuous damage on the intestinal barrier, leading to a progressive and lasting increase in gut permeability. As shown in Fig.~\ref{fig:stress_scenarios}(c), this sustained permeability results in chronic endotoxin leakage, establishing a state of persistent, low-grade systemic inflammation due to elevated endotoxin and TNF-$\alpha$.

This inflammatory signal that feeds back to the brain via the gut-to-brain pathway. This feedback route, as conceptualized in Fig.~\ref{fig:feedback_mc}, becomes the dominant driver of HPA axis activity, overriding the canonical negative feedback mechanism. Unlike in the acute scenario, the system fails to recover and instead enters the pathological state. It becomes trapped in a self-reinforcing loop where elevated cortisol damages the gut barrier and leads to more inflammation that drives cortisol even higher. This demonstrates how a chronic stressor can push the GBA network past a tipping point, locking it into a diseased state characterized by a complete loss of circadian rhythm and a flat profile of hypercortisolism.

\subsection{Bifurcation Analysis and Tipping Points}

To investigate the system's resilience to stress, a one-parameter bifurcation analysis was performed to identify the critical thresholds at which the GBA network transitions from healthy to pathological state. The analysis involved systematically sweeping the constant external stress input, represented by the $k_{\text{leak}}$, across a range of values. For each stress level, the stability of the resulting cortisol oscillation was quantified to map the response of the end-to-end communication system.

\begin{figure}[H]
	\centering
	\includegraphics[width=1\columnwidth]{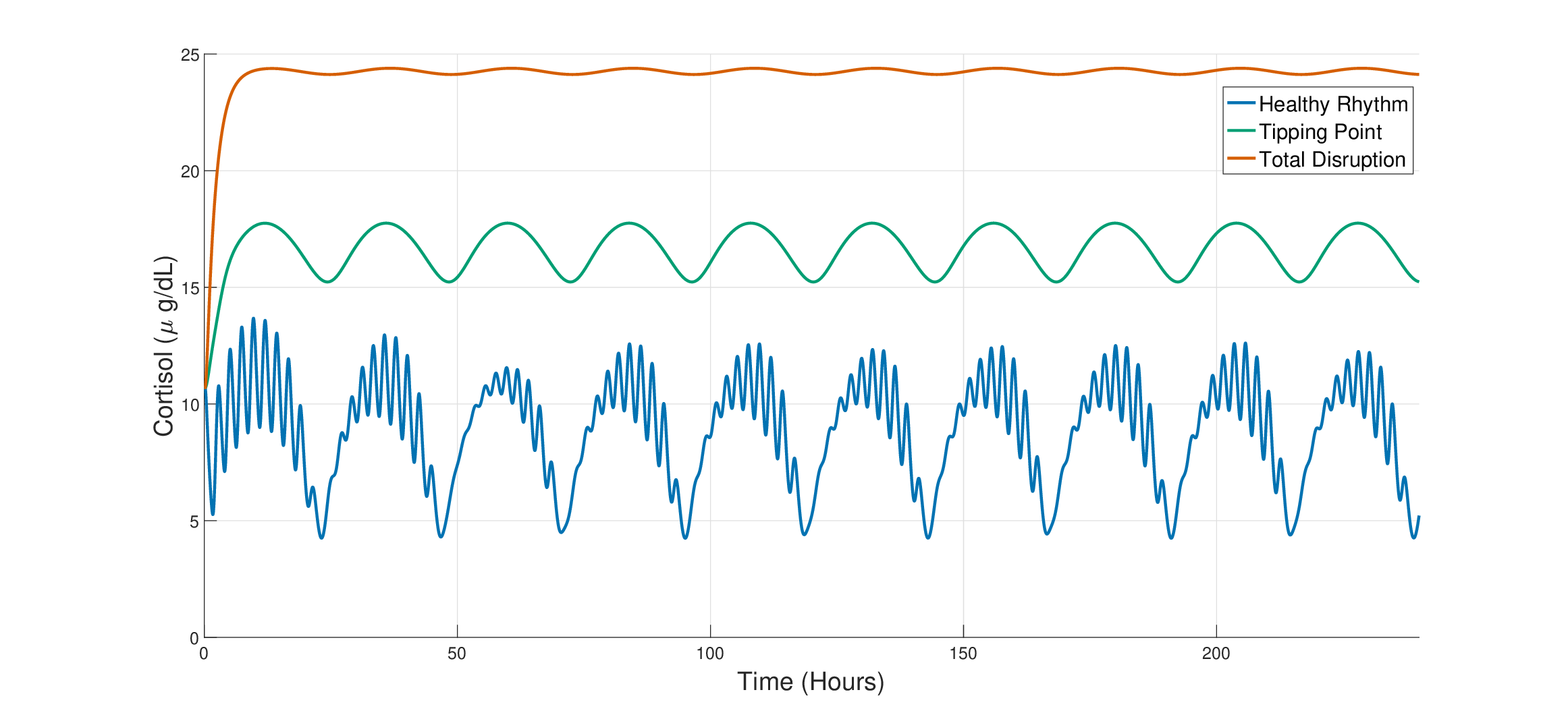}
    \caption{Results of the numerical bifurcation sweep identifying critical thresholds for system stability.}
	\label{fig:cortisol_dynamics}
\end{figure}

The results of this analysis are summarized in Fig.~\ref{fig:cortisol_dynamics}, which compares the system's steady-state cortisol dynamics at three key identified points.

\begin{enumerate}
    \item \textit{Healthy Rhythm:} In the absence of external stress, the system exhibits robust, stable circadian oscillations. This represents a healthy, homeostatic state where the HPA axis's natural negative feedback loop is dominant. This state corresponds to a resilient channel capable of efficiently processing signals.

    \item \textit{The Tipping Point:} As the stress level increases, the system reaches its first critical threshold at a stress input leading $k_{\text{leak}} \approx 1.52$ min$^{-1}$. At this tipping point, the pathological positive feedback from the gut-to-brain inflammatory pathway becomes strong enough to compete with the HPA axis's regulatory negative feedback, significantly dampening the cortisol rhythm and marking the onset of a dysregulated state.

    \item \textit{Total Disruption:} A further increase in stress pushes the system past a second threshold that represents $k_{\text{leak}} \approx 2.04$ min$^{-1}$. After this point, the self-perpetuating inflammatory feedback loop completely overpowers the system's natural regulation, and the circadian oscillation ceases entirely. The system becomes locked in a stable pathological state of hypercortisolism, which corresponds to a state of communication failure with a severely compromised ability to process information.

\end{enumerate}

This bifurcation analysis, therefore, reveals that the shift from health state to pathology is not gradual but occurs at distinct tipping points. This denotes a fundamental change in the system's dynamics. The following section provides a detailed frequency-domain analysis to deepen the understanding of the changes in the channel properties that drive this resilience loss.

\subsection{Frequency Domain Analysis}
To further characterize the GBA as a communication channel, the time-domain simulations are complemented by a frequency-domain analysis. Since the system model described in Sec.~\ref{sec:system_model} is nonlinear and includes explicit delays, a small-signal analysis is performed. In this method, the system is linearized about a stable operating point, from which a transfer function characterizing the response to small perturbations is derived. For this analysis, the six-state DDE system is defined in the following compact state-space form:

\begin{equation}
\begin{split}
    \dot x(t) & = f\!\big(x(t),\,x(t-\tau_{\text{hpa}}),\,x(t-\tau_{\text{gut}}),\,u(t)\big) \\
    y(t) & =C(t) ,
\end{split}
\label{eq:compact_dde_revised}
\end{equation}

\noindent where the state vector $x(t)$ and its time derivative $\dot{x}(t)$ are column vectors defined as:

\begin{equation}
    x(t) = 
    \begin{bmatrix}
        P(t) \\ T(t) \\ S(t) \\ A(t) \\ C(t) \\ L(t)
    \end{bmatrix}, \qquad
    \dot{x}(t) = 
    \begin{bmatrix}
        dP(t)/dt \\ dT(t)/dt \\ dS(t)/dt \\ dA(t)/dt \\ dC(t)/dt \\ dL(t)/dt 
    \end{bmatrix}.
\label{eq:x_openForm}
\end{equation}

Within this state-space framework, the input is defined as $u(t) \triangleq k_{\text{leak}}(t)$, and the output $y(t)$ is the cortisol concentration, selected via the vector $\mathbf{C}_{\text{out}}=[0,0,0,0,1,0]$. The function $f$ represents the complete set of six coupled differential equations from Section~\ref{sec:system_model}. To linearize the system, a stationary operating point, $x^\star$, must first be found. This equilibrium state is determined by fixing the time-varying circadian drive at its mean, $E(t)\equiv\bar{E}$, setting the input to a constant baseline value, $u^\star$, and numerically solving the algebraic system $f(x^\star,x^\star,x^\star,u^\star) = 0$.

The nonlinear system is then approximated by a linear model that is valid for small perturbations, $\delta x(t)=x(t)-x^\star$, around this equilibrium. A first-order Taylor expansion yields the linear model:

\begin{equation}
\label{eq:lindde_revised}
\begin{split}
    \delta\dot x(t) &=\mathbf{J}_0\,\delta x(t) +\mathbf{J}_{\text{hpa}}\,\delta x(t-\tau_{\text{hpa}})  \\ \\
    &+\mathbf{J}_{\text{gut}}\,\delta x(t-\tau_{\text{gut}}) +\mathbf{B}\,\delta u(t),
\end{split}
\end{equation}

The system matrices are the Jacobians, which are composed of the partial derivatives of the system function $f$ evaluated at the equilibrium point ($|_\star$). The instantaneous Jacobian, $\mathbf{J}_0$, is a 6x6 matrix describing the immediate influence of each state variable on every other:

\begin{equation}
    \mathbf{J}_0 = \left.\frac{\partial f}{\partial x}\right|_\star =
    \begin{bmatrix}
        \frac{\partial(dP/dt)}{\partial P} & \cdots & \frac{\partial(dP/dt)}{\partial L} \\
        \vdots & \ddots & \vdots \\
        \frac{\partial(dL/dt)}{\partial P} & \cdots & \frac{\partial(dL/dt)}{\partial L}
    \end{bmatrix}_\star .
\end{equation}

The matrices representing the delayed feedback effects are sparse, as they isolate only the specific delayed interactions. The matrix $\mathbf{J}_{\text{hpa}}$ captures the delayed HPA axis feedback. Since only the equations for $dA/dt$ and $dC/dt$ depend on delayed cortisol and ACTH respectively, its form is:

\begin{equation}
    \mathbf{J}_{\text{hpa}} = \left.\frac{\partial f}{\partial x_{\text{hpa}}}\right|_\star =
    \begin{bmatrix}
        0 & 0 & 0 & 0 & 0 & 0 \\
        0 & 0 & 0 & 0 & 0 & 0 \\
        0 & 0 & 0 & 0 & 0 & 0 \\
        0 & 0 & 0 & 0 & \frac{\partial(dA/dt)}{\partial C(t-\tau_{\text{hpa}})} & 0 \\
        0 & 0 & 0 & \frac{\partial(dC/dt)}{\partial A(t-\tau_{\text{hpa}})} & 0 & 0 \\
        0 & 0 & 0 & 0 & 0 & 0 
    \end{bmatrix}_\star . 
\end{equation}

Similarly, $\mathbf{J}_{\text{gut}}$ captures only the delayed effect of cortisol on gut permeability:

\begin{equation}
    \mathbf{J}_{\text{gut}} = \left.\frac{\partial f}{\partial x_{\text{gut}}}\right|_\star =
    \begin{bmatrix}
        0 & 0 & 0 & 0 & 0 & 0 \\
        0 & 0 & 0 & 0 & 0 & 0 \\
        0 & 0 & 0 & 0 & 0 & 0 \\
        0 & 0 & 0 & 0 & 0 & 0 \\
        0 & 0 & 0 & 0 & 0 & 0 \\
        0 & 0 & 0 & 0 & \frac{\partial(dL/dt)}{\partial C(t-\tau_{\text{gut}})} & 0
    \end{bmatrix}_\star . \\
\end{equation}

The input matrix $\mathbf{B}$ represents how the input perturbation $u(t)$ affects the system's states Since $u(t)$ only appears in the equation for $dP/dt$, the $\mathbf{B}$ is obtained as:

\begin{equation}
    \mathbf{B} = \left.\frac{\partial f}{\partial u}\right|_\star = 
    \begin{bmatrix} \partial(dP/dt)/\partial u \\ 0 \\ 0 \\ 0 \\ 0 \\ 0 \end{bmatrix}_\star =
    \begin{bmatrix} L^\star \\ 0 \\ 0 \\ 0 \\ 0 \\ 0 \end{bmatrix}.
\end{equation}

Note that the linearization of the product term $k_{\text{leak}}L(t)$ from the endotoxin equation yields two distinct components. The first, which enters the input matrix $\mathbf{B}$, shows that the input's effect is scaled by the baseline permeability $L^\star$. The second, which is added to the instantaneous Jacobian matrix $\mathbf{J}_0$, reflects that the baseline leakage rate $u^\star$ modulates the coupling between permeability and endotoxin.

Applying the Laplace Transform to the linearized model \eqref{eq:lindde_revised} converts the system into the following algebraic equation in the frequency domain ($s=j\omega$):

\begin{equation}
    sX(s)=\big(\mathbf{J}_0+\mathbf{J}_{\text{hpa}}e^{-s\tau_{\text{hpa}}}+\mathbf{J}_{\text{gut}}e^{-s\tau_{\text{gut}}}\big)X(s)+\mathbf{B}\,U(s).
\end{equation}

The output of the end-to-end communication channel is the cortisol concentration, $Y(s)$, which is selected from the state vector using the vector $\mathbf{C}_{\text{out}}=[0,0,0,0,1,0]$. The end-to-end transfer function, $H(s)$, is obtained by taking the ratio of this specific output to the input, $H(s) = Y(s)/U(s)$:

\begin{equation}
\label{eq:frf_main}
    H(s) = \mathbf{C}_{\text{out}}\!\left(s\mathbf{I} - \mathbf{J}_0 - \mathbf{J}_{\text{hpa}}e^{-s\tau_{\text{hpa}}} - \mathbf{J}_{\text{gut}}e^{-s\tau_{\text{gut}}}\right)^{-1}\!\mathbf{B} .
\end{equation}

The transfer function in \eqref{eq:frf_main} provides a complete input-to-output description of the linearized system. In this expression, the vector $\mathbf{C}_{\text{out}}$ selects cortisol as the output, and $\mathbf{B}$ maps the input signal to its initial state. The matrix $\mathbf{J}_0$ captures the instantaneous system dynamics, while the exponential terms containing $\mathbf{J}_{\text{hpa}}$ and $\mathbf{J}_{\text{gut}}$ incorporate the crucial feedback delays.

\begin{figure}[H]
	\centering
	\includegraphics[width=\columnwidth]{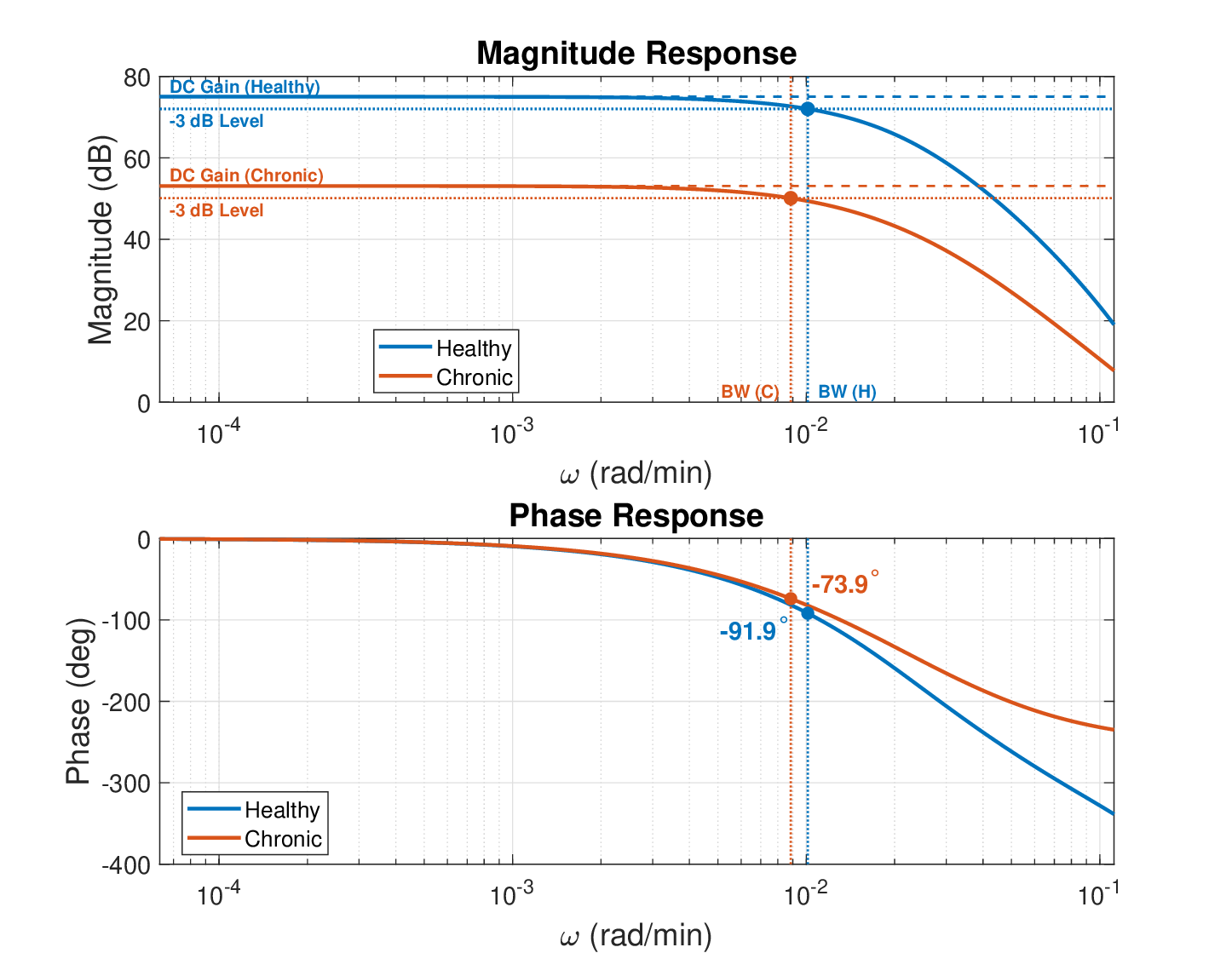}
    \caption{Magnitude and phase response of the linearized GBA channel.}
	\label{fig:bode_healthy_chronic}
\end{figure}

The frequency response of the GBA channel, obtained by evaluating the transfer function at $s=j\omega$, reveals its core signal-processing characteristics. As illustrated in Fig.~\ref{fig:bode_healthy_chronic}, the channel behaves as a high-order low-pass filter. This behavior is governed by the slow biochemical elimination and repair kinetics, which enable the system to naturally attenuate rapid, high-frequency fluctuations. The explicit time delays in the model introduce a substantial, frequency-dependent phase lag, a critical factor that governs the system's closed-loop stability. Since the model is linearized around a fixed point by freezing the circadian input $E(t)$ at its mean, it does not capture the dominant 24-hour resonance of the full nonlinear system. A more rigorous treatment would require periodic (Floquet) linearization. It is important to note that this linearization operation is valid only in the vicinity of a locally asymptotically stable equilibrium point, $x^\star$. Outside this stable regime, the frequency response is not predictive.

\subsection{Quantitative Performance Metrics: DC Gain and Bandwidth}
The transfer function, \eqref{eq:frf_main}, provides a basis for quantifying the channel's performance using standard metrics. Two key metrics, the static (DC) gain and the bandwidth, are used to summarize the end-to-end behavior of the linearized GBA channel.

The DC gain is found by evaluating the transfer function at zero frequency:

\begin{equation}
\label{eq:static_dc}
    H(0)\;=\;-\,\mathbf{C}_{\text{out}}\big(\mathbf{J}_0+\mathbf{J}_{\text{hpa}}+\mathbf{J}_{\text{gut}}\big)^{-1}\mathbf{B}.
\end{equation}

A large magnitude of $|H(0)|$ indicates strong amplification of a persistent change in gut leakage into a steady offset in cortisol. As the system's operating point approaches a bifurcation threshold (a tipping point), the total system matrix, $\mathbf{J}_\Sigma \triangleq \mathbf{J}_0+\mathbf{J}_{\text{hpa}}+\mathbf{J}_{\text{gut}}$, becomes ill-conditioned, causing $|H(0)|$ to grow significantly. The DC gain therefore serves as an indicator for the system's proximity to a loss of stability.

The responsiveness of the channel is characterized by its bandwidth. The half-power bandwidth, $\omega_{3\text{dB}}$, is a standard metric defined as the frequency at which the channel's gain drops to $1/\sqrt{2}$ (or -3 dB) of its DC value:

\begin{equation}
\label{eq:3db}
    |H(j\omega_{3\text{dB}})|=\frac{|H(0)|}{\sqrt{2}}.
\end{equation}

The DC gain $|H(0)|$ quantifies the steady-state sensitivity to persistent changes in $k_{\text{leak}}$, whereas the half-power bandwidth $\omega_{3\mathrm{dB}}$ summarizes the channel’s responsiveness to changing inputs. As illustrated in Fig.~\ref{fig:bode_healthy_chronic}, the healthy operating point exhibits a broader effective passband than the chronic state, consistent with the bandwidth narrowing expected near a loss of resilience. These two metrics provide the basis for interpreting the information-theoretic results that follow.

\subsection{Information-Theoretic Channel Capacity}

The transfer function provides a basis for calculating the channel's information-theoretic capacity, which defines the maximum rate of reliable information transmission for small perturbations. To perform this calculation, the linearized GBA channel is modeled as a continuous-time, single-input single-output (SISO) system with additive noise:

\begin{equation}
    y(t)=\big(h * u\big)(t)+n(t),
\end{equation}

where $h(t)$ is the impulse response corresponding to the transfer function $H(s)$, $u(t)$ is the input perturbation to $k_{\text{leak}}$, and $n(t)$ is a zero-mean, stationary Gaussian noise process with power spectral density (PSD) $S_n(\omega)$. The input signal is constrained by an average power budget, $P_{av}$:

\begin{equation}
    \frac{1}{2\pi}\int_{-\infty}^{\infty} S_u(\omega)\,d\omega \le P_{av},
\end{equation}

For a stable LTI channel with wide-sense stationary Gaussian noise, a power-constrained Gaussian input achieves the Shannon capacity, which is defined as \cite{cover1999elements}:

\begin{equation}
\label{eq:ct_capacity}
    C = \frac{1}{2\pi}\int_{-\infty}^{\infty}
    \log_2\!\left(1+\frac{|H(j\omega)|^2\,S_u^\star(\omega)}{S_n(\omega)}\right)\,d\omega,
\end{equation}

\noindent where $S_u^\star(\omega)$ is the optimal input power spectrum that maximizes the capacity. This optimal spectrum is found using the "water-filling" algorithm, which strategically allocates more power to frequencies where the channel gain is high and the noise is low:

\begin{equation}
\label{eq:water_filling}
    S_u^\star(\omega)\;=\;\max\left\{0, \;\mu-\frac{S_n(\omega)}{|H(j\omega)|^2}\right\},
\end{equation}

\noindent where the constant $\mu$ (the “water level”) is chosen to satisfy the total power constraint, $\displaystyle \frac{1}{2\pi}\!\int_{-\infty}^{\infty}\! S_u^\star(\omega)\,d\omega=P_{\text{av}}$.

The water-filling algorithm allocates input power strategically to maximize information flow. It assigns more power to frequencies with a higher signal-to-noise ratio (SNR) density, given by the term $|H(j\omega)|^{2}/S_n(\omega)$. For a low-pass channel such as the GBA, this generally concentrates power within the effective passband. However, the allocation is not a hard cutoff at a specific frequency but smoothly follows the entire SNR profile. A crucial property of this formula is that capacity depends on the channel's magnitude response, $|H(j\omega)|$, and the noise PSD. The channel's phase response does not affect the total capacity. In our model, the physiological time delays ($\tau_{\text{hpa}}, \tau_{\text{gut}}$) are critical because they indirectly alter the capacity by reshaping the magnitude response, as captured in the transfer function \eqref{eq:frf_main}.

\begin{figure}[H]
	\centering
	\includegraphics[width=0.9\columnwidth]{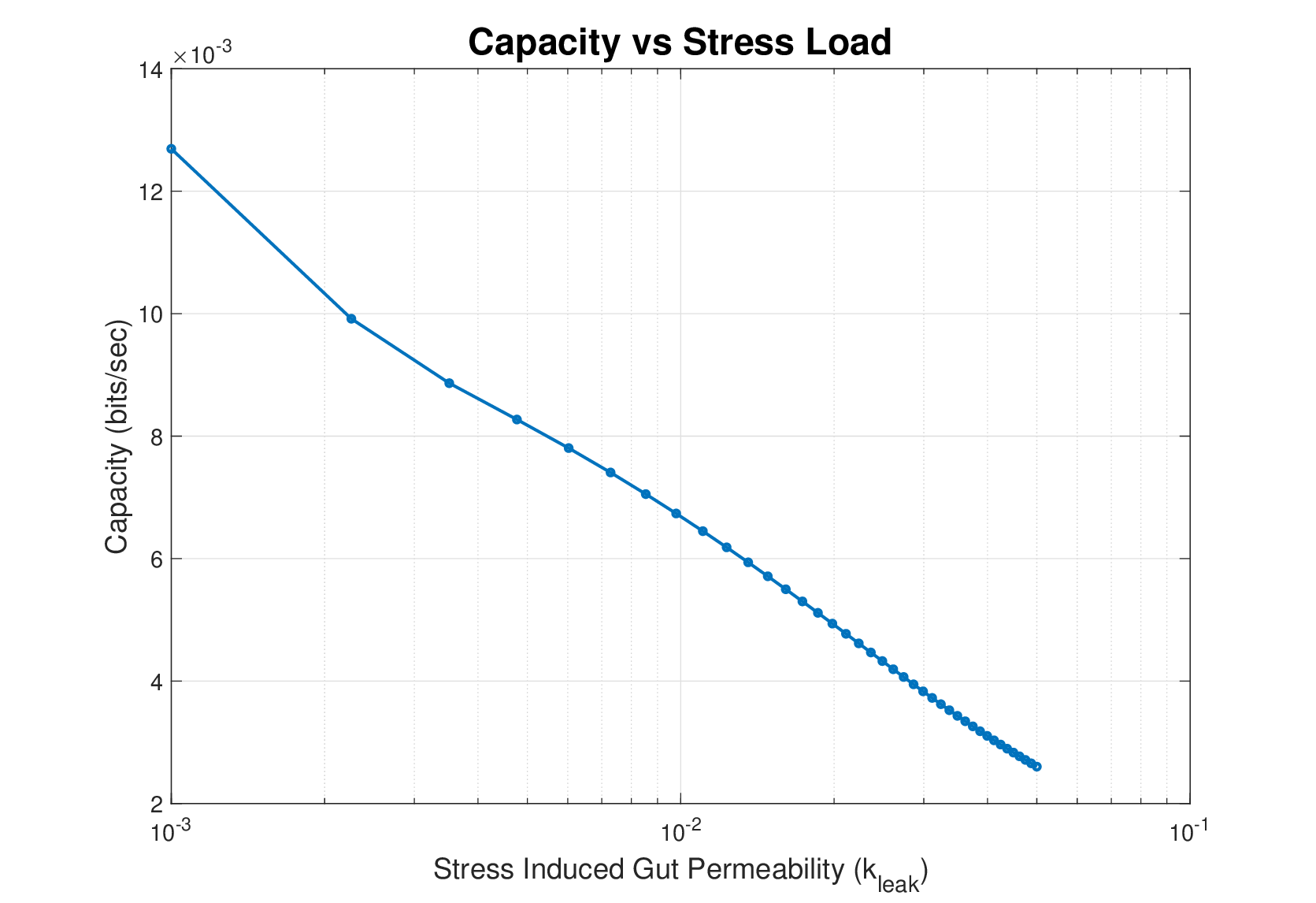}
    \caption{Information-theoretic channel capacity as a function of stress-induced leakage rate ($k_{\text{leak}}$).}	
    \label{fig:stress_vs_capacity}
\end{figure}

The capacity in \eqref{eq:ct_capacity} is a small-signal upper bound for the linearized, stable operating point with Gaussian noise and an average-power constraint. The achievable rate in a real biological context would be lower due to factors such as non-Gaussian disturbances and physiological amplitude constraints on the signaling molecules. As the transfer function $H(s)$ depends on the equilibrium, the capacity is also operating-point dependent. As shown in Fig. \ref{fig:stress_vs_capacity}, this capacity varies as the system transitions from healthy to stressed states, revealing that increasing chronic stress levels progressively degrade the channel's ability to transmit information reliably.

\begin{figure*}[t]
	\centering
	\includegraphics[width=\textwidth]{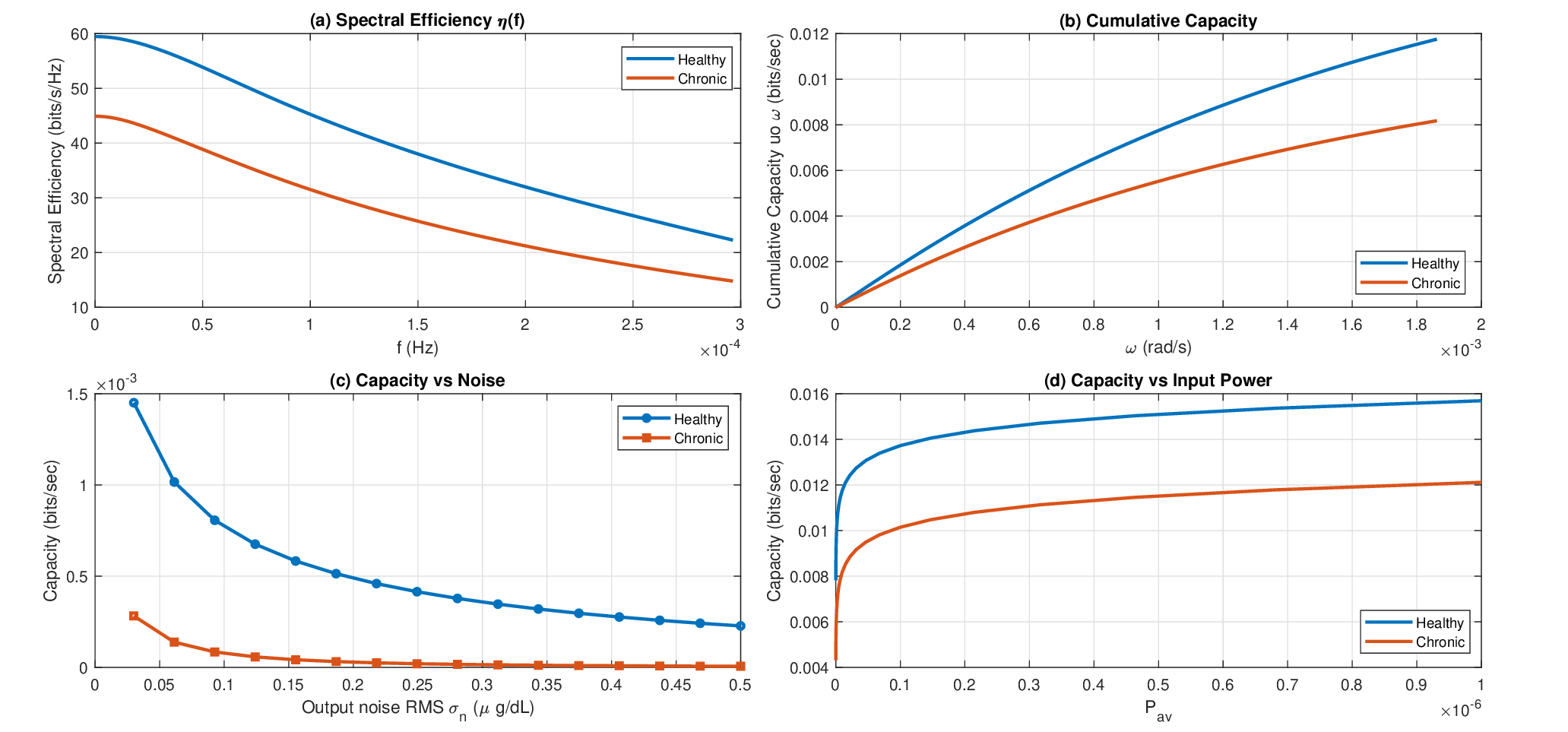}
    \caption{Detailed analysis of channel capacity for healthy and chronic stress states.}
	\label{fig:capacity_subplot}
\end{figure*}

To further examine the communication impairment, Fig. \ref{fig:capacity_subplot} provides a detailed comparison between the healthy and chronic stress operating points. In Fig.~\ref{fig:capacity_subplot} (a), the spectral efficiency $\eta(f)$ (bits/s/Hz) is concentrated at low frequencies $f$, and the healthy state exhibits a higher peak and a broader effective passband, consistent with a wider usable bandwidth. In Fig.~\ref{fig:capacity_subplot}(b), the cumulative capacity $C(\omega)=\int_{0}^{\omega}\eta\!\big(\omega'/(2\pi)\big)\,\tfrac{d\omega'}{2\pi}$ (with $\omega$ in rad/s) rises more rapidly and reaches a larger value in the healthy state, confirming greater total information throughput. In Fig.~\ref{fig:capacity_subplot} (c), capacity decreases monotonically as the output noise level increases, while the healthy state preserves a consistent advantage that indicates superior robustness. In Fig.~\ref{fig:capacity_subplot} (d), capacity increases concavely with the input power budget and shows diminishing returns at high power. The healthy state achieves higher rates for any fixed power, demonstrating better power efficiency. Collectively, the results quantify the communication performance deficit under chronic stress since chronic stress squeezes the usable spectrum, reduces passband gain, and limits the information processing capability of the GBA channel. The outcomes demonstrate that the pathological state is characterized not only by hormonal dysregulation but also by a fundamental impairment in the ability to process and transmit information reliably and efficiently.

\section{Discussion} 
\label{sec:discussion}

This work introduces a novel molecular communication framework that models the gut-brain axis (GBA) as a closed-loop system governed by delayed feedback mechanisms. The proposed model, which integrates gut barrier dynamics with the canonical HPA axis and peripheral immune signaling, enables a quantitative analysis of how bidirectional communication governs the body's response to stress for changing conditions. The time-domain simulations reveal the GBA's two distinct operational regimes. In the healthy state, HPA axis-driven negative feedback dominates, enabling the stable circadian cortisol rhythms observed in Fig.~\ref{fig:stress_scenarios}(a). The system demonstrates rapid recovery from acute stressors (Fig.~\ref{fig:stress_scenarios}(b) and Fig.~\ref{fig:acute_inf}) since the end-to-end channel behaves as a high-order low-pass filter around its homeostatic operating point. This low-pass characteristic, which arises from the slow biochemical kinetics and explicit time delays, allows the system to attenuate transient, high-frequency perturbations effectively. However, a sustained stress input ignites a pathological positive feedback loop where cortisol-induced gut damage drives inflammation, which in turn stimulates more cortisol release. This self-reinforcing cycle leads to the collapse of circadian oscillations shown in Fig.~\ref{fig:stress_scenarios}(c). Although the delay-induced phase is negligible at very low frequencies, it accumulates with frequency and progressively erodes the effectiveness of negative feedback. This explains why short, high-frequency perturbations are attenuated while prolonged inputs can shift the operating point and unlock the self-reinforcing loop.

The transition from a healthy to a pathological state is not gradual but occurs at critical thresholds identified as tipping points in the bifurcation analysis of Fig.~\ref{fig:cortisol_dynamics}. As the system approaches these thresholds, it exhibits a clear signature of critical slowing down, resulting in a significant narrowing of its communication bandwidth (Fig.~\ref{fig:bode_healthy_chronic}). This narrowing mathematically corresponds to the dominant eigenvalues of the system's Jacobian matrix approaching the imaginary axis, indicating longer recovery times. While the approach to a bifurcation is typically marked by increasing static gain, comparison of the final stable equilibria reveals that the DC gain is substantially lower in the chronic state than in the healthy one. This occurs because the chronic high-cortisol equilibrium drives the system into a saturated regime. At this high operating point, the nonlinear Hill functions in the model desensitize the feedback loops, reducing the system's local linear sensitivity to small perturbations. Therefore, the pathological state is characterized by a communication channel that is both less responsive, as shown by the narrower bandwidth, and less sensitive to persistent inputs, as reflected by the reduced DC gain.

An information-theoretic analysis translates these dynamic changes into quantitative performance metrics. The chronic stress state is characterized by a compressed usable spectrum, reduced total throughput, and lower resilience to noise, as detailed in Fig.~\ref{fig:stress_vs_capacity} and Fig.~\ref{fig:capacity_subplot}. This degradation in communication performance directly results from the altered channel transfer function, $H(s)$, in the chronic state. Channel capacity is fundamentally determined by integrating the spectral efficiency, which is a function of the signal-to-noise ratio (SNR), across the usable frequency spectrum. As shown in Fig.~\ref{fig:bode_healthy_chronic}, the chronic equilibrium results in a channel with a lower passband gain ($|H|$) and a narrower bandwidth. These changes directly reduce the SNR over the frequencies, which lowers the spectral efficiency. When integrated, this diminished spectral efficiency results in a lower total channel capacity, as confirmed by the water-filling evaluation. This analysis demonstrates that the pathology involves not just hormonal dysregulation but a fundamental degradation of the GBA's ability to transmit information reliably. This suggests that communication metrics such as channel capacity could serve as holistic biomarkers of GBA health.

The simulation outcomes validate the proposed end-to-end closed-loop communication model. The stable circadian rhythms in the healthy baseline confirm the proper implementation of the HPA axis's dominant negative feedback loop. The system's ability to recover from acute stress validates the low-pass filter characteristic created by the model's slow biochemical kinetics, which correctly attenuates transient perturbations. Finally, the simulated collapse into a chronic state of hypercortisolism confirms that the model accurately captures the pathological positive feedback loop that emerges when sustained stress engages the system's nonlinearities and time delays. These observations show that the proposed framework successfully maps core physiological mechanisms to their expected dynamic behaviors.

The proposed framework also illustrates potential therapeutic targets for restoring healthy communication. The results suggest that interventions aimed at strengthening gut barrier integrity, enhancing the clearance of inflammatory molecules, or weakening the stimulatory coupling between cytokines and the HPA axis are all expected to increase system resilience. This modeling approach thus enables the in-silico exploration of treatments designed to shift the system's operating point away from the pathological region and restore robust information flow within the GBA.

\section{Conclusion} 
\label{sec:conclusion}

This paper introduces a novel, end-to-end molecular communication framework which models the gut-brain axis as a fully integrated, closed-loop system. The model is formulated as a system of six coupled delay differential equations that explicitly unites three critical physiological subsystems, namely the integrity of the gut barrier, the dynamics of pro-inflammatory cytokines, and the neuroendocrine signaling of the HPA axis. This unified structure provides a powerful tool for investigating how bidirectional communication between the gut and brain governs the body's response to stress.

The developed communication framework uncovers the fundamental mechanism of stress-induced pathology. Quantitative analysis demonstrates that prolonged stress drives the GBA past a tipping point, shifting its dynamics from a stable state of negative feedback to a pathological one dominated by a self-perpetuating positive feedback loop. This study quantifies this state transition using communication theory, establishing a direct link between the loss of physiological stability and a severe degradation in channel performance. The pathological state is characterized by a reduced bandwidth, lower information capacity, and decreased robustness to molecular noise. This framework provides a novel understanding that GBA dysregulation is a fundamental impairment of the body's information processing, rather than just a hormonal imbalance.

This comprehensive model enables the uncovering of communication mechanisms within the GBA, paving the way for unprecedented applications to explore diverse scenarios and communication pathways. By systematically linking physiological conditions to information processing metrics such as passband gain, effective bandwidth, noise resilience, and channel capacity, this framework enables the in-silico screening of therapeutic interventions. This communication-centric approach promises to re-establish reliable information flow, offering new possibilities for treating stress-related disorders. Future work will focus on validating these predictions with clinical data and extending the model to include other key signaling pathways for patient-specific personalization.

\bibliographystyle{IEEEtran}
\bibliography{References}

\begin{IEEEbiography}
    [{\includegraphics[width=1in, height=1.2in, clip, keepaspectratio]{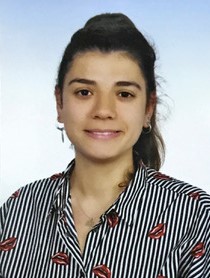}}]{Beyza Ezgi Ortlek} (Graduate Student Member, IEEE) received the B.Sc. degree in electrical and electronics engineering and in molecular biology and genetics as a double major from Koç University, Istanbul, Türkiye, where she is currently pursuing the Ph.D. degree in electrical and electronics engineering and also a Research Assistant with the Center for neXt-generation Communications (CXC). Her research interests include intrabody nanonetworks and molecular communications.
\end{IEEEbiography}

\begin{IEEEbiography}
[{\includegraphics[width=1in,height=1.2in,clip,keepaspectratio]{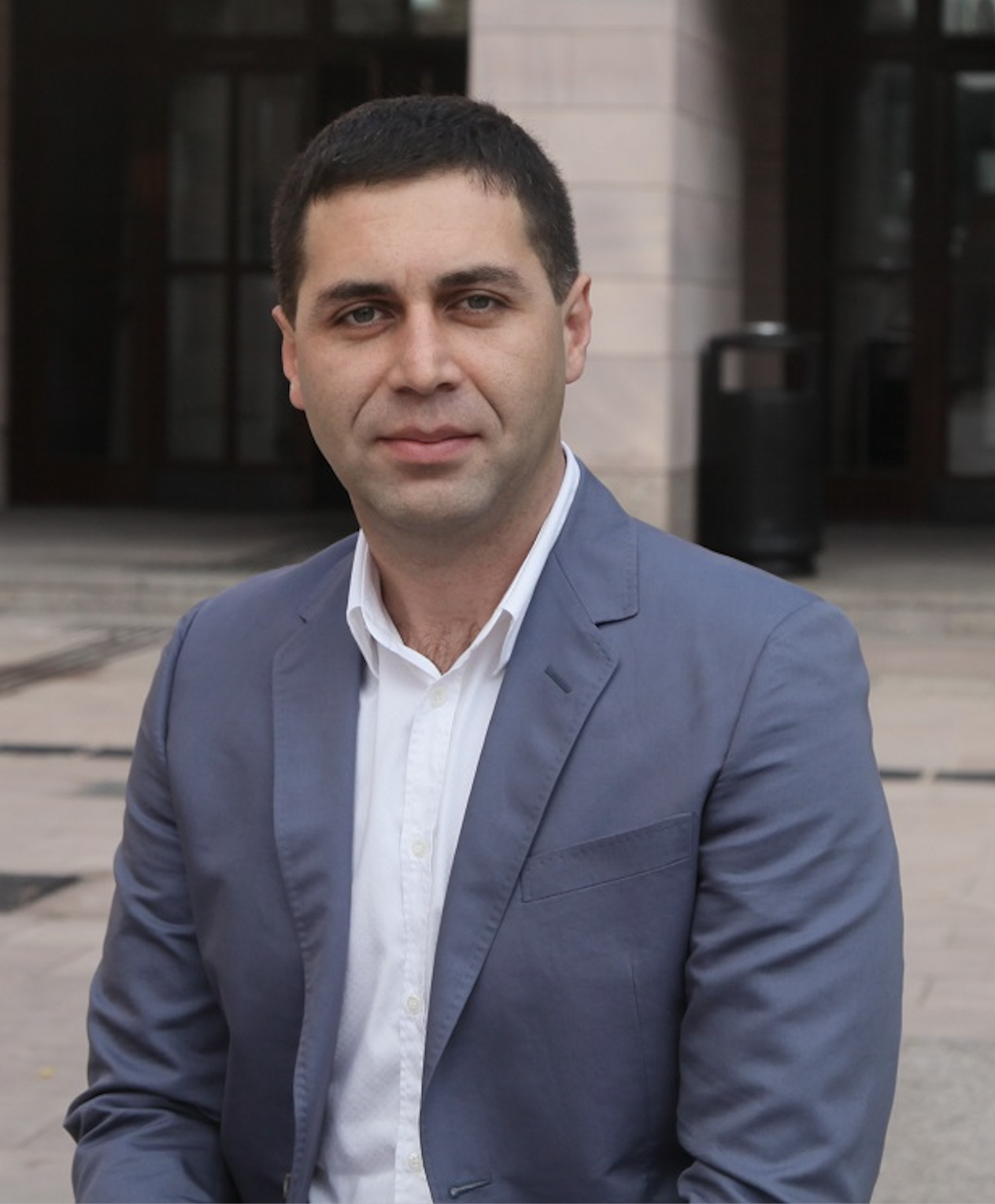}}]{Ozgur B. Akan} (Fellow, IEEE)
received the PhD from the School of Electrical and Computer Engineering Georgia Institute of Technology Atlanta, in 2004. He is currently the Head of Internet of Everything (IoE) Group, with the Department of Engineering, University of Cambridge, UK and the Director of Centre for neXt-generation Communications (CXC), Koç University, Turkey. His research interests include wireless, nano, and molecular communications and Internet of Everything.
\end{IEEEbiography}

\end{document}